# GraphHP: A Hybrid Platform for Iterative Graph Processing

Qun Chen, Song Bai, Zhanhuai Li, Zhiying Gou, Bo Suo and Wei Pan
*Northwestern Polytechnical University*
Xi'an, China
{chenbenben, baisong, lizhh, gzhiying, suobo, panwei}@nwpu.edu.cn

## ABSTRACT

The Bulk Synchronous Parallel (BSP) computational model has emerged as the dominant distributed framework to build large-scale iterative graph processing systems. While its implementations (e.g., Pregel, Giraph and Hama) achieve high scalability, frequent synchronization and communication among the workers can cause substantial parallel inefficiency. To help address this critical concern, this paper introduces the *GraphHP* (*Graph H*ybrid *P*rocessing) platform which inherits the friendly vertex-centric BSP programming interface and optimizes its synchronization and communication overhead.

To achieve the goal, we first propose a hybrid execution model which differentiates between the computations within a graph partition and across the partitions, and decouples the computations within a partition from distributed synchronization and communication. By implementing the computations within a partition by pseudo-superstep iteration in memory, the hybrid execution model can effectively reduce synchronization and communication overhead while not requiring heavy scheduling overhead or graph-centric sequential algorithms. We then demonstrate how the hybrid execution model can be easily implemented within the BSP abstraction to preserve its simple programming interface. Finally, we evaluate our implementation of the GraphHP platform on classical BSP applications and show that it performs significantly better than the state-of-the-art BSP implementations. Our GraphHP implementation is based on Hama, but can easily generalize to other BSP platforms.

## 1. INTRODUCTION

An increasing number of big data applications are focused on computations over graphs modeling the *dependencies* between data: examples include Web and social network analysis [8, 31], graph mining [33], and machine learning [14, 24]. For the problems just cited, as well as a number of others, a common property of the analysis is a sequence of iterative computations over data dependencies. With the exponential growth in the scale of these problems, there is an increasing need for systems that can execute the analysis efficiently in parallel on large clusters.

Designing and implementing large-scale distributed processing systems can be prohibitively challenging because it requires programmers to address deadlocks, data race, distributed state and communication protocols. To avoid repeatedly solving the same distributed computing problems, high-level parallel programming models (e.g., MapReduce [13, 2] and Dryad [18]) have been proposed to insulate programmers from the complexities of distributed implementation. Unfortunately, the parallel platform like MapReduce was not designed for iterative computation, thus not well suited for dependent analysis. While it is often possible to map dependent computations into the MapReduce abstraction, the resulting transformations may be challenging, introducing suboptimal performance and usability issues.

The limitations of the MapReduce abstraction have driven the community to build the *vertex-centric* parallel platforms (e.g., Pregel [23], Giraph [1] and Hama [3]) based on the Bulk Synchronous Parallel (BSP) model [34]. The computations on the BSP platforms consist of a sequence of iterations, called *supersteps*. The BSP approach is reminiscent of MapReduce in that users define a local action for each superstep, processing each vertex independently, and the system composes these actions to lift computations to a large graph. Nonetheless, the BSP approach better fits iterative graph algorithms and makes it easier to reason about program semantics while implementing algorithms. Unfortunately, implementing efficient graph algorithms on Pregel-like systems can be challenging, primarily due to slow convergence and huge communication cost [28, 27].

The existing work to address the inefficiencies of the BSP platforms can be classified into two categories. One, including distributed GraphLab [22] and Giraph++ [32], deviates from the BSP programming model. GraphLab adopts an asynchronous computational model allowing users to directly read and modify the values at adjacent vertices. Besides the locking mechanisms to enforce data consistency, it also requires heavy scheduling overhead. Giraph++ advocates a graph-centric programming interface requiring users to program complex sequential algorithms for graph partitions. Its efficiency depends greatly on the user-defined sequential algorithms. The other consists of the piecemeal solutions [28, 27, 7, 10] that complement the BSP systems. Even though these proposed techniques can reduce synchronization and communication overhead, they are either tailored to specific graph algorithms, thus having limited usability, or marginal optimizations that do not change the underlying BSP execution model, thus having limited effectiveness. The examples of the first case include the Storing Edges At Subvertices (SEAS) and Single Pivot (SP) methods [28], which can be applied to the minimum spanning forest and connected component problems respectively. The method of implementing message passing between vertices within a worker directly in memory [23, 7] is an example of the other case. Some of the proposals (e.g., Finishing Computation Serially (FCS) [27] even require a trade-off between synchronization and communication as well as a not-so-simple extended API.

With the increasing use of BSP platforms, there is an urgent need for a general platform that can address the inefficiency concern on BSP synchronization and communication while at the same time preserving its friendly vertex-centric programming interface. To fill this critical void, we introduce the general-purpose distributed platform GraphHP that can significantly reduce synchronization and communication overhead without sacrificing the simple BSP programming interface. To achieve this goal, we propose a hybrid execution model that differentiates between the computations within a graph partition and across partitions, and decouples the computations within a partition from distributed communication and synchronization. By implementing the computations within a partition by pseudo-superstep iteration in memory, the hybrid execution model can effectively reduce synchronization and communication overhead while not requiring heavy scheduling overhead. We provide a formal description of the hybrid execution model and demonstrate how it can be easily implemented within the BSP abstraction. Our main contributions are as follows:

- We analyze the performance of the standard BSP platforms and summarize their limitations in implementing efficient iterative graph algorithms, making the case for a general platform that preserves the BSP programming interface while being able to optimize synchronization and communication overhead.

- We propose a hybrid execution model that compared with the standard BSP execution model, has the potential to achieve significantly better parallel performance by reducing the frequency of global iterations.

- We design and implement the hybrid platform GraphHP for iterative graph processing. GraphHP inherits the simple vertex-centric BSP programming interface, but has a distinct hybrid execution model. Our implementation is based on Hama, but can easily generalizes to other BSP platforms.

- We evaluate the performance of GraphHP on classical BSP applications by comparative study. Our comprehensive experiments demonstrate that GraphHP achieves significantly better performance than the state-of-the-art BSP implementations. It also has potential performance advantage over the asynchronous platform GraphLab and the graph-centric platform Giraph++.

The rest of this paper is organized as follows: Section 2 analyzes the performance of the existing BSP platforms. Section 3 describes the BSP programming interface. Section 4 introduces the hybrid GraphHP execution model. Section 5 presents the design and implementation of the GraphHP platform. Section 6 discusses the application of GraphHP on classical BSP computations. Section 7 presents our empirical evaluation results. Section 8 reviews related work. Finally, Section 9 concludes this paper with some thoughts on future work.

## 2. ANALYSIS OF BSP PLATFORMS

The popularity of the BSP model, evidenced by many BSP implementations (e.g., Pregel [23], Hama [3] and Giraph [1]), mainly arises from its scalability and its flexible and easy to use "think like a vertex" programming interface. The interface naturally fits the dependent computations of iterative graph algorithms centered on graph vertices. The synchronicity of the BSP model liberates programmers from the burden of specifying order of execution within an iteration and also ensures that its programs are inherently free of deadlocks and data race common in asynchronous systems. In principle, the performance of BSP programs should be competitive with that of asynchronous systems given enough parallel slack.

Unfortunately, the increasing use of the BSP platforms has also exposed their limitations. It was observed in [28] that even implementing standard graph algorithms (e.g., strongly connected components, minimum spanning forest and graph coloring) can incur substantial inefficiency of slow convergence due to structural properties of the input graph. Making things worse, many machine learning and data mining algorithms (e.g., brief propagation [14] and stochastic optimization [30]) have inherently slow convergence rates, which make their efficient and scalable implementations on the BSP platforms even more challenging.

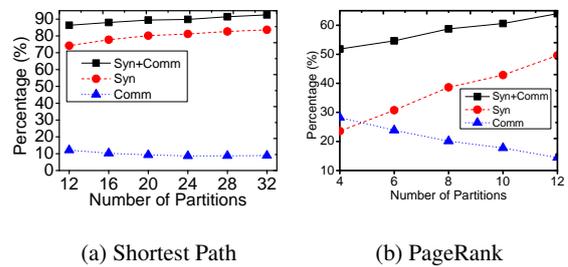

(a) Shortest Path  (b) PageRank

**Figure 1: Synchronization and Communication Overhead**

To dissect the cost model of BSP programs, we empirically study the performance of the shortest path and incremental PageRank BSP algorithms [23, 36] on the platform Hama on a ten-machine cluster. They are evaluated on a USA road network data (USA-Road-NE) from [5] and a Web graph data (Web-Google) from [4] respectively. We plot the synchronization and communication overhead, as a percentage of the whole processing cost against the number of partitions in Figure 1. The cost is measured by the elapsed time. It is averaged over all the participating workers. As shown in Figure 1 (a), for the shortest path computation, synchronization and communication combined accounts for the whopping 86% of the whole processing time even with only 12 partitions. The overhead percentage consistently increases as the number of partitions increases. Synchronization solely accounts for the overwhelming 74% of the whole processing cost with only 12 partitions. The percentage also consistently increases as the number of partitions increases. In contrast, the communication overhead measured by percentage consistently decreases as the number of partitions increases. For the PageRank computation, as shown in Figure 1 (b), the BSP platform performs in a similar way. Synchronization and communication combined account for the majority of the processing time. These experiments demonstrate that instead of the execution on vertices, synchronization and communication dominate the computations of BSP programs. Both synchronization and communication can contribute substantially to the inefficiency of BSP programs, with synchronization usually playing a more significant role.

Unsurprisingly, many optimization techniques [23, 7, 28, 27] have been developed to reduce the synchronization and communication overhead. The typical BSP platforms like Pregel usually provide the functionality of *combiner*, which can combine several messages intended for a vertex into a single message, to reduce communication overhead. As pointed out in [7], communication overhead can be further reduced by combining all the messages intended for a worker and letting the workers deliver the messages to their destination vertices. This approach is similar to the *combiner* functionality in that it combines many replicas of the same message intended for the same worker into a single one. [32] also proposed a hybrid communication mechanism which performs message passing between vertices within a same partition directly in memory while executing message passing across partitions in the same way as typical BSP platforms. On synchronization overhead, [28] introduced a Finishing Computations Serially (FCS) method that transfers the task of processing an active small graph with slow convergence rate to the master. The master performs the computation on the small graph serially and then sends the results back to the worker. While these proposals are effective in practice if applicable, they are piecemeal solutions that address specific inefficiencies present in BSP programs and usually require users to provide with optimization details.

The limitations of the existing BSP platforms and their optimization techniques motivate us to develop the platform GraphHP with a hybrid execution model in this paper. GraphHP is designed with the assumption that the attractive properties of standard BSP platforms, which include the simple vertex-centric programming interface and bulk vertex processing without heavy scheduling overhead, should be preserved. GraphHP can significantly reduce the synchronization and communication inefficiencies common to iterative graph processing while not requiring specific optimization instructions from users.

## 3. BSP PROGRAMMING INTERFACE

The BSP data model is a directed graph in which each vertex is uniquely identified by a vertex identifier. The directed edges are associated with source vertices, whose neighbors are target vertices. Both vertex and edge have updatable states but only vertex has its associated computation. The BSP computation consists of a sequence of supersteps. During a superstep, the framework invokes a uniform, user-defined function for each vertex, conceptually in parallel. The function specifies the behavior at a single vertex $v$ and a single superstep ($S$). It can read messages sent to $v$ in superstep ($S$-$1$), send messages to other vertices that will be received at superstep ($S+1$), and update the states of the vertex $v$ and its outgoing edges.

We illustrate the BSP programming interface by Hama, an open-source BSP implementation. The programming interface of Hama mainly consists of the following classes and methods:

- The `Vertex` class. This is the most important class responsible for instructing the behaviors of vertices and edges and maintaining their states. Its primary method `Compute()` defines the actions taken at each active vertex in every superstep. `Compute()` can inspect the received messages via a message iterator and send messages using the method `sendMessage()`. It can query and update the state of a vertex using the methods `getValue()` and `setValue()` respectively.

- The `Aggregator` class. It is a mechanism for global communication and monitoring. Each vertex can submit a value to an aggregator in superstep ($S$). The aggregator reduces the received values into a single one and makes it available to all vertices in superstep ($S+1$). Typical operations provided by the aggregator include `min`, `max` and `sum`.

- The `Combiner` class. It is a mechanism to reduce communication overhead by combining several messages intended for a vertex into a single one. Enabling this optimization requires the user to specify a combination rule in the virtual method `Combine()`.

---

**Algorithm 1:** A PageRank Implementation: `Compute()`

// *Msg* denotes the incoming message queue for the vertex $v$;
// *numV* denotes the total number of vertices in the input graph;
// *N(v)* denotes the set of $v$'s neighboring vertices;
**if** *getSuperstepCount()==0* **then**
    setValue($\frac{1}{numV}$);
**else**
    setValue($\frac{0.15}{numV}$ + 0.85×sum(Msg)) ;
**if** *(getSuperstepCount()<30)* **then**
    sendMessageToNeighbors($\frac{getValue()}{|N(v)|}$);
**else**
    voteToHalt();

---

Writing a BSP program involves subclassing the predefined `Vertex` class. The user defines three value types associated with vertices, edges and messages, and overrides the virtual `Compute()` method. The user can also subclass the `Combiner` and `Aggregator` classes to implement the message combination and value aggregation functions on vertices. The pseudo-code of the `Compute()` function for a straightforward PageRank implementation, which uses the real interface provided by Hama, is shown in Algorithm 1.

## 4. HYBRID EXECUTION MODEL

BSP programs are executed on a machine cluster consisting of a master and multiple workers. The master is not assigned any portion of the input graph, but is responsible for coordinating worker activity. Each worker is assigned one or more partitions, and is responsible for all the activities on its section of the graph. In this section, we first present the standard BSP execution model and then introduce the hybrid execution model of GraphHP.

### 4.1 Standard Model

The foremost characteristic of the standard execution model is uniformity. Each vertex has a computational state, *active* or *inactive*. The execution consists of a sequence of iterations. Each iteration corresponds to a superstep, in which every *active* vertex performs the operations defined by the single `Compute()` function. Even though the function behaviors may vary, the serial actions at a vertex $v$ during a superstep $S$ generally consist of the following steps: (1) retrieve the messages sent to $v$ at superstep ($S$-$1$); (2) perform the user-defined operations on $v$ and its outgoing edges; (3) send messages to other vertices that will be processed at superstep ($S+1$); (4) update its computational state if necessary; (5) flag its computational state.

The standard model controls algorithm progress through the mechanism of vertex self-deactivation. Initially, the computational state of every vertex is set to be *active*. All the active vertices participate in the computations of any given superstep. During its computation, a vertex may choose to deactivate itself. This means that the vertex will not participate in the computations of subsequent supersteps unless it receives a message, automatically reactivating itself. If reactivated, a vertex must explicitly deactivate itself again. An algorithm terminates if and only if all the vertices are simultaneously *inactive* and no message is in transit.

The communication between vertices is achieved through asynchronous message passing. By default, messages are transferred through distributed communication mechanisms like RPC (Remote Procedure Call). In the case that the target vertex is located at the same worker as the source vertex, message passing can instead be directly performed in memory. Between superstep iterations, every message should be delivered to its destination vertex. The vertices should also inform the worker of their states, which are then made available to the master. Finally, the master synchronizes all the workers and instructs them to proceed to the next superstep simultaneously.

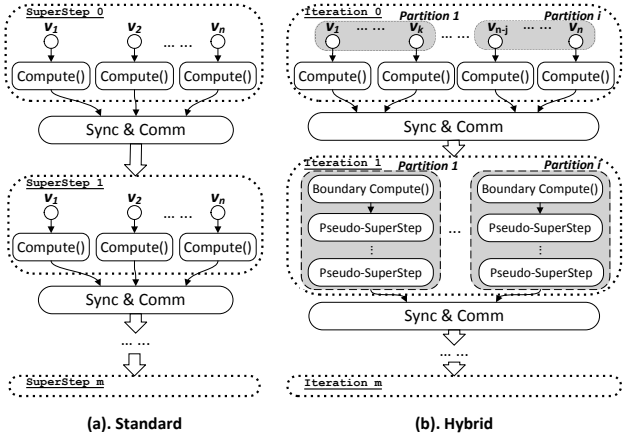

**Figure 2: Execution Models: Standard vs Hybrid**

The standard execution model is abstracted in Figure 2 (a). Between superstep iterations, it has to perform synchronization and message passing, which are usually even more time-consuming than the computation itself (as observed in Section 2). Synchronous computation also incurs performance penalty since the run time of each superstep is determined by its slowest worker. The penalty may become costly if accumulated over lots of supersteps. In the case of slow convergence, frequent superstep invocation, triggering equally frequent communication and synchronization, can seriously limit the efficiency of BSP programs. This motivates us to develop the hybrid execution model of GraphHP whose most important purpose is to effectively reduce global iteration invocation frequency.

### 4.2 Hybrid Model

Before describing the hybrid model in details, we first give some denotations to simplify the presentation.

DEFINITION 1. (Local Vertex and Boundary Vertex). *Within a graph partition, a vertex $v$ is called a* local vertex *if all the source vertices of its incoming edges are located in the same partition as $v$. Otherwise, $v$ has at least an incoming edge whose source vertex is located in a remote partition other than $v$'s; it is called a* boundary vertex.

DEFINITION 2. (Local Computation and Boundary Computation). *The* Compute() *operation at a local vertex is called* local computation. *The operation at a boundary vertex is called* boundary computation.

It is obvious that local computation does not need to directly communicate with other partitions. In contrast, boundary computation involves remote communication across the partitions. The hybrid model considers each graph partition as a computational unit and processes local and boundary computations separately. It consists of a sequence of global iterations, each including a global phase and a local phase. The global and local phases are responsible for boundary and local computations respectively. By default, boundary vertices participate in the computations of the global phase while the local phase only involves the computations on the local vertices.

A global phase is similar to a superstep in the standard model in that each active boundary vertex executes its Compute() function, receiving and sending out messages. A boundary vertex may send out new messages to local and boundary vertices during the execution of a global phase. If intended for a local vertex, the message will be processed at the immediate local phase in the same iteration. The message intended for a boundary vertex will instead be temporarily buffered and processed in the global phase of the next iteration. A message sent to a local or boundary vertex will reactivate the vertex, making its computational state *active*, if it is originally inactive. Since boundary vertices can only receive their most recently sent messages from other partitions at the next iteration, the global phase processes each active boundary vertex only once.

A global phase is followed by a local phase. Conceptually similar to the BSP abstraction of the standard model, a local phase consists of a sequence of pseudo-supersteps. The high-level behaviors of the vertices at pseudo-supersteps are also the same as those defined for the vertices in the standard model. The pseudo-superstep execution is nonetheless different from the standard superstep execution in two aspects. Firstly, each partition processes its pseudo-superstep iteration independently without involving synchronization and communication across the partitions. Secondly, the communication between local vertices is achieved in memory through direct message passing. Local phase termination is achieved through the same vertex self-deactivation mechanism of the standard model. For each partition, its local phase terminates if and only if all of its local vertices become inactive and no message is in transit between them.

Similar to boundary vertices in a global phase, active local vertices may send out new messages during a pseudo-superstep. If intended for a local vertex, the message will be immediately processed in the next pseudo-superstep. The message intended for a boundary vertex will instead be temporarily buffered, and only processed in the global phase of the next iteration. All the messages intended for boundary vertices, sent out in the global and local phase of an iteration, should be delivered to their destination vertices before the beginning of the next iteration. The hybrid model therefore requires distributed synchronization and communication only once for each global iteration. The global iteration is repeatedly executed until the algorithm terminates. An algorithm terminates if and only if all the vertices, including both local and boundary vertices, become inactive simultaneously and no message is in transit.

The hybrid execution model is abstracted in Figure 2 (b). Its first iteration (iteration 0) is an initialization iteration in which vertices are activated, assigned initial values and send out new messages. The hybrid model executes its first iteration in the same way as the standard model executes its first superstep. From the iteration 1 on, it repeatedly invokes a global phase and a local phase. At a global phase, it executes the Compute() function on each active boundary vertex once, using the messages sent to it during the previous iteration as input. This ensures that each boundary vertex is processed with the most recent information about its neighbors.

In the case of incremental computations (e.g., shortest path computation), the Compute() function can be executed on a vertex even if it only receives a portion of the incoming messages intended for it. For instance, in the single-source shortest path computation, the value of a vertex can be safely modified by any incoming mes-sage if it provides a shorter distance value. In this case, boundary vertices can participate in the computations of a local phase without sacrificing algorithmic correctness. The computations on them are executed in the same way as those on local vertices. Our GraphHP implementation allows the user to configure whether or not boundary vertices can participate in the computations of local phases. Since boundary vertices' participation in local phases usually accelerates algorithmic convergence, this feature should be activated whenever applicable.

The hybrid execution model inherits the underlying communication and synchronization mechanisms of the standard BSP execution engine. It achieves reduced global iteration frequency by automatically desynchronizes the computations on the vertices based on their graph partitions. Unlike the asynchronous platform GraphLab, the hybrid model still processes the vertices in *bulk* mode. It does not require heavy scheduling overhead. The hybrid model is also different from the execution engine of Giraph++. Giraph++ requires users to program complex sequential algorithms for graph partitions. These algorithms usually sequentially scan active vertices within a partition and perform corresponding actions. In contrast, the hybrid execution model preserves the uniform and easy to reason vertex-centric BSP programming interface: the computation at any vertex, whether it is local or boundary, is defined by a single `Compute()` function. Within each global iteration, GraphHP iteratively processes active local vertices within a partition until they all become inactive.

We also note that Grace [35], a single-machine parallel platform, proposed an asynchronous messaging mechanism. During the execution of a superstep, a message sent to a vertex in a same partition can be processed by the receiver at the same superstep. Suppose that the vertices $u$ and $v$ are in the same partition, and during a superstep, $u$ sends $v$ a message. The message will be directly put into $v$'s incoming message queue. If later $v$ is processed at the same superstep, it will be immediately used to perform $v$'s `Compute()` function. Otherwise, if $v$ is already been processed before the message is sent, the message will be processed at the next superstep. Its multiple-machine version has also been implemented in [32] for comparative empirical study. The hybrid model is different from the Grace-like platforms in that it repeatedly processes local vertices within a partition until they converge during the local phase of an iteration, while Grace processes each vertex at most once for each iteration.

It is worthy to point out that the hybrid model can take advantage of the asynchronous messaging mechanism to optimize its performance. During the execution of a local phase, a message sent to a vertex in a same partition can be processed by its receiver at the same pseudo-superstep. We have implemented the BSP platform with the asynchronous messaging mechanism on Hama, denoted by AM-Hama, and compared its performance with that of GraphHP in Section 7.

## 5. PLATFORM IMPLEMENTATION

Inheriting the user-friendly vertex-centric BSP programming interface, GraphHP complements it with additional methods to facilitate hybrid execution. GraphHP fully supports the standard BSP classes and methods presented in Section 3. In spite of the hybrid execution model, the behaviors of local and boundary vertices are defined by the same `Compute()` function in the `Vertex` class. Communication between vertices is achieved through message passing, which is specified by the `SendMessage()` method in the `Vertex` class.

As on the standard BSP platforms, writing a GraphHP program involves subclassing the predefined `Vertex` class. The users can specify whether or not boundary vertices will participate in the computations of local phases. Between iterations, a boundary vertex may receive multiple messages from another vertex. Obviously, the user-defined `Combine()` function can be applied to combine these messages. The `Combine()` function, however, is supposed to combine all the messages intended for a vertex. GraphHP provides an additional function, `SourceCombine()`, to combine the messages intended for a vertex and originating from a same source vertex. By default, only the *latest* message is saved. Alternatively, users can manually define any appropriate combination rule.

We built the GraphHP platform based on the standard BSP platform Hama. Involving only minor system adjustments, the GraphHP implementation does not require redesigning the distributed architecture of Hama as well as its underlying communication and synchronization mechanisms. Instead of providing every detail of the GraphHP implementation, we will focus on the mechanisms that enable hybrid execution in the rest of this section. Even though GraphHP is built on Hama, its implementation can easily generalize to other BSP platforms.

### 5.1 Basic Architecture

The GraphHP architecture consists of a master and multiple workers. The master assigns one or more partitions of the input graph to each worker. Besides the unique *id*, each vertex within a graph partition has an indicator indicating a local or boundary vertex. The outgoing edges are represented by the adjacency lists of source vertices. Each vertex in an adjacency list has a location indicator indicating it is located at the same partition as the source vertex or at another partition.

GraphHP executes its initialization iteration in the same way as Hama executes its initialization superstep. After the first iteration, the master instructs each worker to repeatedly execute the global iteration consisting of a global phase and a local phase. The worker uses one thread for each partition to execute both the global and local phases during an iteration. The global phase loops through active boundary vertices and executes the `Compute()` function on each of them. The local phase iteratively invokes a pseudo-superstep. At each pseudo-superstep, the thread loops through active local vertices and executes the `Compute()` function on each of them.

### 5.2 Worker Implementation

GraphHP implements the global phase based on the superstep mechanism of Hama. When `Compute()` requests sending a message to another vertex, the worker first determines whether or not the receiver is located at the same partition as the sender. If yes, the message is directly placed in the destination vertex's incoming message queue. Otherwise, the message will be temporarily buffered and later delivered through RPC as on Hama. Since the messages transferred across the partitions will be processed in the next iteration, they are only required to be delivered before the beginning of the next iteration. When a partition finishes a global phase, it immediately proceeds to a local phase without the need to notify the master of the switch.

Conceptually, GraphHP implements the pseudo-superstep iteration of a local phase in a similar way as Hama implements the superstep iteration. During a pseudo-superstep, the thread loops through the active local vertices and executes the `Compute()` function on each of them. While a worker processes an active local vertex, it may simultaneously send new messages to other vertices. As in the global phase, a message is either directly put into its destination vertex's message queue or temporarily buffered for later delivery depending on the location of its receiver. Executing the pseudo-superstep iteration does not depend on the delivery of buffered messages. Instead, GraphHP only needs to deliver the buffered messages between global iterations. Except asynchronous transferring of the buffered messages, the pseudo-superstep iteration is performed entirely in memory. It does not require communication and synchronization across graph partitions.

The `bsp()` function at the workers is sketched in Algorithm 2, in which a peer corresponds to a graph partition, and *bMsgs* and *lMsgs* represent the message iterators for boundary vertices and local vertices respectively. The *globalSuperstep()* function executes a global phase. The *pseudoSuperstep()* function executes pseudo-supersteps. The `sendMessage()` function is also sketched in Algorithm 3. If the destination vertex is located at a remote peer (partition), the message is inserted into the remote message iterator *rMsgs*, whose contents will be delivered through RPC. Otherwise, if it is a boundary vertex located at the same partition, the message is directly inserted into *bMsgs*. Or if it is a local vertex, the message is directly inserted into *lMsgs*.

---

**Algorithm 2:** The `bsp()` Function

// *aLV* denotes the number of active local vertices;
**while** *algorithmNotFinished* **do**
    peer.syn();
    bMsgs =+ parseMessage(peer);
    masterUpdate(peer);
    globalSuperstep(bMsgs);
    **while** *(aLV!=0) || (lMsgs!=NULL)* **do**
        aLV=pseudoSuperstep(lMsgs);

---

**Algorithm 3:** The `sendMessage(destV,msgs)` Function

// *msgs* denotes the message;
// *destV* denotes the destination vertex of the message;
**if** *peer(destV)!=peer(this)* **then**
    rMsgs =+ msgs;
**else if** *boundary(destV)* **then**
    bMsgs =+ msgs;
**else**
    lMsgs =+ msgs;

---

In case that boundary vertices participate in the computations of local phases, GraphHP processes boundary vertices in the same way as it processes local vertices. Instead of looping through the local vertices, the *pseudoSuperstep()* function will loops through all the active vertices within a partition. In the `sendMessage()` function, all the messages intended for the vertices in the local partition will be inserted into *lMsgs*. As a result, during a local phase, an active boundary vertex will process all the messages from the participating vertices within its own partition, but not the messages from other partitions. Within a partition, its local phase terminates if and only if all of its participating vertices, including both the local and boundary vertices, become inactive and no message is in transit between them.

### 5.3 Master Implementation

The master primarily coordinates the activities of workers. It is responsible for orchestrating the simultaneous iterations on the workers. GraphHP uses the superstep index of Hama as its global iteration index and instructs the workers to simultaneously iterate over the process consisting of a global phase and a local phase.

Note that GraphHP does not need to modify the underlying synchronization mechanism of Hama. It is achieved through *barriers*. The master sends the same request to every worker at the beginning of each iteration, and waits for a response from every worker. If the barrier synchronization succeeds, the master instructs the workers to proceed to the next iteration simultaneously.

GraphHP also inherits the fault tolerance mechanism of Hama. It is implemented through checkpointing. At the beginning of a global or local phase, the master instructs the workers to save the states of their partitions to a permanent storage. In case that local computations within a partition are intensive, GraphHP can opt to enact multiple checkpoints during the execution of a local phase. The master issues regular "ping" messages to workers. If the master does not hear back from a worker within a specified interval, it marks that worker as failed. When a worker fails, the master reassigns its graph partitions to another currently available worker. The new worker will reload their partition states from the most recent checkpoint.

## 6. CASE STUDIES

In this section, we apply GraphHP on three classical BSP applications, shortest path [12], PageRank [8] and bipartite matching [6], and compare their executions with those on Hama. The shortest path and PageRank computations are the typical algorithms of graph traversal and random walk respectively. The bipartite matching computation represents the category of graph analysis algorithms that require different types of messages to be sent and processed at different stages of the computation.

### 6.1 Shortest Paths

The shortest paths problem is one of the best known graph problems. For the comparative purpose, we focus here on the single-source shortest path problem (SSP) which searches for the shortest path between a single source vertex and every other vertex in a graph.

---

**Algorithm 4:** The `Compute()` Function for SSP

// *Msg* denotes the message queue for the vertex $v$;
// $N(v)$ denotes the set of $v$'s neighboring vertices;
// $d(v,u)$ denotes the distance from the vertex $v$ to $u$;
**if** *getSuperstepCount()==0* **then**
    **if** *source(v)* **then**
        setValue(0);
        **for** $u \in N(v)$ **do**
            sendMessage($u$, getValue()+$d(v,u)$);
    **else**
        setValue($\infty$);
**else**
    newValue=min(Msg);
    **if** *newValue<getValue()* **then**
        setValue(newValue);
        **for** $u \in N(v)$ **do**
            sendMessage($u$, newValue+$d(v,u)$);
voteToHalt();

---

The pseudo-code of an implementation on Hama is shown in Algorithm 4. Initially, the value of the source vertex is set to 0 while the values of other vertices are all set to $\infty$ (a constant larger than any feasible distance). The source vertex also propagates its value to its immediate neighbors. These neighbors in turn will update their values and send messages to their neighbors, resulting in a wavefront of updates through the graph. On Hama, a superstep can only propagate the values one vertex away. Since every vertex is only interested in the shortest distance, it will update its value if and only if it receives a message containing a smaller one. A Combine() function can be specified to combine the messages intended for a vertex into a single one containing the smallest value.

The Hama implementation can be reused on GraphHP. Boundary vertices can participate in the computations of local phases. After the first initialization iteration, GraphHP iteratively executes a global phase followed by a local phase. The global phase will propagate the updated values of boundary vertices across partitions. The local phase iteratively propagates the updates of boundary vertices to all the vertices within a partition until the value of every vertex becomes stable. The specified Combine() function on Hama can also be used to combine all the messages intended for a same vertex. The hybrid execution model of GraphHP can significantly reduce the required global iteration frequency. Please refer to Section 7.2 for detailed experimental results.

## 6.2 PageRank

---
**Algorithm 5:** The Compute() Function for Incremental PageRank

---
// $\Delta$ is the user-defined convergence tolerance;
**if** *getSuperstepCount()==0* **then**
    setValue(0);
    updateValue=0.15;
**else**
    updateValue=sum(Msg);
**if** *updateValue* $> \Delta$ **then**
    setValue(getValue()+updateValue);
    **for** $u \in N(v)$ **do**
        sendMessage(u, $\frac{updateValue}{|N(v)|}$);
voteToHalt();

---

A straightforward PageRank implementation, as shown in Algorithm 1, iteratively updates a vertex value based on the values from the previous superstep. It requires the vertices to remain active and continue sending messages even after their computations have converged. Otherwise, the vertices may fail to receive the necessary values from the previous superstep since some vertices may have converged and stop propagating their values. To avoid redundant message passing, we can alternatively implement an accumulative iterative update BSP algorithm [36], whose pseudo-code is shown in Algorithm 5. The incremental algorithm accumulates the intermediate updates to an existing PageRank value. During each superstep, an active vertex will propagate its PageRank value update to its immediate neighbors. A Combine() function can be specified to sum up the value updates intended for a same vertex. The superstep is repeatedly invoked until the value of every vertex converges within a predefined tolerance.

The implementation of the incremental algorithm on Hama can also be reused on GraphHP. For the incremental algorithm, boundary vertices can participate in the computations of local phases. After the initialization iteration, GraphHP begins the second iteration with a global superstep, in which each partition updates the PageRank values of its boundary vertices. In the following local phase, the participating vertices, including both local and boundary vertices, iteratively update their PageRank values through pseudo-supersteps until their values all converge. The iteration is repeatedly invoked until all the vertices become inactive and no message is in transit, which mean every vertex's PageRank value has converged. Between iterations, if a vertex sends multiple messages to a same destination vertex, the user-defined Combine() function can be applied to sum up their value updates before delivery. GraphHP effectively encapsulates the convergence computations within a partition in a local phase. It can effectively reduce the frequency of global synchronization and communication. Please refer to Section 7.3 for detailed experimental results.

## 6.3 Bipartite Matching

A bipartite graph consists of two distinct sets of vertices with edges only between the sets. A bipartite matching is a subset of edges without common endpoints. The bipartite matching (BM) problem is to find the maximal matchings in which no additional edge can be added without sharing an end point.

---
**Algorithm 6:** The Compute() Function for Bipartite Matching on GraphHP

---
// *Msg* denotes the message queue for the vertex $v$;
// *vid(msgs)* denotes the sender of the message *msgs*;
// *state(v)* denotes the algorithmic state of the vertex $v$;
**if** *left(v)* **then**
    **if** *Msg==NULL* **then**
        sendMessageToNeighbors("request");
        voteToHalt();
    **else**
        **while** *(msgs=Msg.getNext())!=NULL* **do**
            **if** *msgs=="grant"* **then**
                **if** *state(v)!=matched* **then**
                      setValue(vid(msgs));
                      state(v)=*matched*;
                      sendMessage(vid(msgs),"accept");
                **else**
                      sendMessage(vid(msgs),"deny");
            voteToHalt();
**else if** *right(v)* **then**
    **if** *state(v)==ungranted* **then**
        rNum=random(Msg.size());
        count=0;
        **while** *(msgs=Msg.getNext())!=NULL* **do**
            **if** *count==rNum* **then**
                state(v)=*granted*;
                sendMessage(vid(msgs), "grant");
            **else**
                sendMessage(vid(msgs), "deny");
            count++;
    **else if** *state(v)==granted* **then**
        **while** *(msgs=Msg.getNext())!=NULL* **do**
            **if** *msgs=="accept"* **then**
                setValue(vid(msgs));
                state(v)=*matched*;
            **else if** *msgs=="deny"* **then**
                state(v)=*ungranted*;
            **else if** *msgs=="request"* **then**
                **if** *state(v)==ungranted* **then**
                      sendMessage(vid(msgs), "grant");
                      state(v)=*granted*;
                **else**
                      sendMessage(vid(msgs), "deny");
voteToHalt();

---

A straightforward BSP implementation on Hama iteratively executes the cycle consisting of four stages. In the 1st stage, each active and unmatched left vertex sends a message to each of its neigh-bors to request match, and then unconditionally votes to halt. In the 2nd stage, each unmatched right vertex randomly chooses one of the received messages, sends a corresponding *grant* message, and sends *deny* messages to other requesters. Finally, it unconditionally votes to halt. In the 3rd stage, each unmatched left vertex chooses one of the grants it receives and send a corresponding acceptance message. If it does not receive any grant or denial, it unconditionally votes to halt. If it receives only denials but no grant, it will remain active and continue to send out new match requests in the 1st stage of the next cycle. In the 4th stage, an unmatched right vertex receives at most one acceptance message. It records its corresponding left vertex and unconditionally votes to halt. In the Hama execution, each of the 2nd and 3rd stages corresponds to a superstep while the 1st and 4th stages can be simultaneously performed at a same superstep. The algorithm effectively terminates when every left vertex either is matched, or has no neighboring right vertex not yet matched.

The bipartite matching BSP algorithm requires the vertices to process different types of messages at different stages of the algorithm. Because of its asynchronous execution model, GraphHP requires a more stringent handshake mechanism to establish the matches between left and right vertices. Moreover, a right vertex may simultaneously receive a search request and an acceptance response from its neighboring left vertices. The GraphHP program, therefore, needs to specify not only the handshake mechanism but the actions upon hybrid message queues. The pseudo-code of the GraphHP implementation is shown in Algorithm 6. Left vertices has two types of algorithmic states, *unmatched* and *matched*. Right vertices have three types of algorithmic states, *ungranted*, *granted* and *matched*. The state of *ungranted* means that a right vertex has not granted any match request yet. In contrast, the state of *granted* means that a right vertex has granted a match request, sending out a *grant* message, but has not received an *accept* message yet. A right vertex in the *granted* state can not grant any new match request, but will send a *deny* message to each requester.

The GraphHP program similarly consists of four stages. In the 1st stage, the unmatched left vertices send out match requests. In the 2nd stage, each ungranted right vertex randomly chooses a match request, and sends out corresponding *grant* and *deny* messages. It also updates its algorithmic state to *granted*. In the 3rd stage, each unmatched left vertex chooses one of the grants it received, sending an *accept* message to the granter, and send a *deny* message to each of other granters. In the 4th stage, if a right vertex in the *granted* state receives an *accept* message, it is matched. Otherwise, if it receives a *deny* message, it will update its algorithmic state back to *ungranted*.

Suppose that boundary vertices participate in the computations of local phases. In the initialization iteration (iteration 0), each left vertex sends out match requests to its neighboring right vertices and then unconditionally votes to halt. In the global phase of the iteration 1, every boundary right vertex with incoming messages performs the actions of the 2nd algorithmic stage, scanning the match requests it received and sends out corresponding *grant* and *deny* messages. In the first pseudo-superstep of the following local phase, every local right vertex with incoming messages similarly performs the actions of the 2nd stage. Simultaneously, each left vertex with incoming messages performs the actions of the 3rd stage. In the following pseudo-supersteps, the algorithmic states of the left and right vertices evolve by the rules programmed in Algorithm 6. The local phase iteratively invokes pseudo-supersteps to execute the four-stage matching process until every left vertex is either matched to a right vertex, or has no neighboring right vertex not yet matched within its own partition.

The *grant* and *deny* messages, sent by the boundary right vertices across the partitions in the iteration 1, will trigger the active states of their receiving left vertices at the beginning of the iteration 2. In the global phase of the iteration 2, each active boundary left vertex performs the actions of the 3rd algorithmic stage. An unmatched boundary left vertex will remain active if it received any *deny* message in the global phase. It will continue to send out new match requests in the following local phase. Since it has no neighboring unmatched right vertex in its own partition, it will not receive any *deny* message during the local phase. The local phase will terminate after at most two pseudo-supersteps. Note that the matching between left and right vertices within a same partition is achieved by a single iteration (the iteration 1). From then on, GraphHP repeatedly invokes global iterations to execute the matching across partitions until a maximal matching is found.

## 6.4 Discussion

By implementing communication between vertices by message passing and limiting the `Compute()` function's data access to a vertex and its outgoing edges, the standard BSP platforms ensure that the scopes of concurrently executing `Compute()` functions (even on adjacent vertices) do not overlap. GraphHP preserves the underlying BSP synchronization and communication implementations as well as the vertex-centric programming interface. It is therefore well suited for the BSP programs in which all the exchanged messages are of a uniform type. For the algorithms like bipartite matching and other computations involving topology mutations, a vertex may simultaneously receive different types of messages during a GraphHP iteration. The BSP programming interface of GraphHP provides the users with the necessary flexibility to specify the required actions on the vertices receiving heterogeneous messages.

## 7. EXPERIMENTAL STUDY

In this section, we empirically evaluate the performance of GraphHP on three classical BSP applications, shortest path, PageRank and bipartite matching. We compare GraphHP with the state-of-the-art BSP platform Hama and its optimized version AM-Hama, which implements the asynchronous messaging mechanism [35, 32] described in Section 4.2. AM-Hama has a similar execution engine as Hama but processes the messages in an asynchronous way. If a message is intended for a vertex located at a remote partition, it is transferred by the distributed mechanism RPC as on Hama. The message will only be processed at the next superstep. Otherwise, it is processed in memory, directly placed into the incoming message queue of its destination vertex. During a superstep, an active vertex will process all the messages in its message queue at the time of its execution. For fair comparison, a same `Combine()` function is specified to combine the messages intended for a same vertex on all the platforms if applicable. On GraphHP, boundary vertices participate in the computations of local phases and the asynchronous messaging mechanism is activated if applicable.

We also compare GraphHP with two other parallel platforms, distributed GraphLab and Giraph++. Unfortunately, we can not conduct a comprehensive head-to-head comparison between GraphHP and either of them. GraphLab is a distributed platform specifically designed to support asynchronous iterative computing. It has a different vertex-centric programming interface allowing users to directly read and modify the values of adjacent vertices. In contrast, GraphHP implements the communication between vertices by message passing. Moreover, GraphLab is written in C++ while GraphHP is written in Java. These implementation details can contribute a great deal to their performance difference.

| Dataset | Algorithm | Size(MB) | $|V|$ | $|E|$ | Description |
|---|---|---|---|---|---|
| USA-Road-NE | SSP | 78.4 | 1,524,453 | 3,897,636 | Northeast USA road network [5] |
| USA-Road-Full | SSP | 325 | 23,947,347 | 58,333,344 | Full USA road network [5] |
| Web-Google | PageRank | 71.8 | 916,428 | 5,105,039 | Web graph from the Google programming contest, 2002 [4] |
| uk-2002 | PageRank | 4690 | 18,520,486 | 298,113,762 | A 2002 crawl of the .uk domain performed by UbiCrawler [4] |
| cit-patents | BM | 249 | 3,774,768 | 16,518,948 | Citation network among US patents [4] |
| delaunay_n24 | BM | 800 | 16,777,216 | 50,331,601 | Delaunay triangulations of random points in the plane [4] |

**Table 1: Details of Test Datasets**

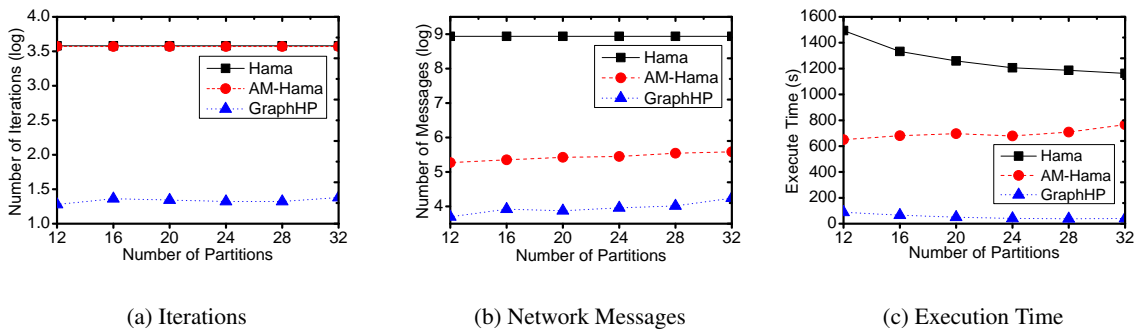

(a) Iterations  (b) Network Messages  (c) Execution Time

**Figure 3: Evaluation Results of SSP on USA-Road-NE**

Giraph++ [32] is a distributed platform with a graph-centric programming interface. It was made clear by the authors that its central contribution is a flexible graph-centric programming model. The performance of Giraph++ depends greatly on the user-defined sequential algorithm for graph partitions. In contrast, GraphHP is designed on the assumption that the friendly vertex-centric BSP programming interface should be preserved. Therefore, our focus is not on their head-to-head performance comparison, but to demonstrate that with the proposed hybrid execution model, GraphHP has the potential to outperform a state-of-the-art asynchronous platform and a graph-centric platform requiring users to write complicated sequential programs.

The rest of this section is organized as follows. In Section 7.1, we describe the experimental setup. In the sections from 7.2 to 7.4, we compare the performance of Hama, AM-Hama and GraphHP on the three BSP applications respectively. Finally, we compare the performance of GraphHP with that of GraphLab and Giraph++ in Section 7.5.

## 7.1 Experimental Setup

The details of the test datasets are summarized in Table 1. The first five datasets are good representatives of real life graphs with the heavy-tail degree distribution. They are often used to evaluate their respective algorithms. The last one, delaunay_n24, is a Delaunay graph widely used in the evaluation of graph partitioning and clustering algorithms. Since maximal matching is a fundamental operation for graph partitioning and clustering, we use the delaunay_n24 dataset for the evaluation of the BM algorithm.

Our machine cluster consists of one master and twelve slaves. Each machine runs the Ubuntu Linux (version 10.04). It has a memory size of 16G, disk storage of 160G and 16 AMD Opteron(TM) processors with the frequency of 2600MHz. They are interconnected using 1Gbit Ethernet.

We implemented GraphHP based on the Hama platform. In its default setting, Hama assigns a vertex to a partition by a hash function ($hash(id) \bmod k$), where $id$ is the vertex identification and $k$ is the number of partitions. Obviously, this random partitioning results in a large number of edges crossing the partition boundaries. A good partitioning strategy should minimize the number of edges connecting different partitions, to potentially reduce communication overhead during a distributed computation. The graph partitioning heuristic Metis [20] is often used to generate better partitions. In the case that the input graph becomes too big to be processed by a single machine, its parallel version ParMetis [19], which implements a parallel multi-level k-way graph partitioning algorithm, can be employed. Alternatively, a variant of ParMetis based on graph coarsening [32] can also be used to partition big graphs. The focus of our experimental study is to evaluate the performance of different distributed platforms. We use ParMetis to divide the test graphs and assign vertices to the resulting partitions accordingly. A study on the quality of different partitioning schemes is beyond the scope of this paper.

Note that running the same algorithm on different platforms has some common overhead cost, such as setting up a job, reading the input into memory, writing the final output to a permanent storage, and shutting down the job. The common overhead difference between different platforms is negligible. In order to focus on the processing efficiency of different platforms, we exclude the common overhead from the reported execution time for all the experiments. All timings are averaged over three or more runs.

## 7.2 Shortest Path

The performance of Hama, AM-Hama and GraphHP are compared on three metrics, the number of global iterations, the number of network messages, and the execution time.

The comparative evaluation results for the SSP algorithm on the USA-Road-NE data are shown in Figure 3. Note that because of huge performance difference, the Y-axis in Figure 3 (a) and (b) is presented by a *logarithmic* scale (base 10). Compared with Hama, AM-Hama can only reduce the numbers of required iterations marginally (from 3800+ to 3700+). In contrast, GraphHP only requires around 20 iterations, reducing them by ratios of hundreds. On number of network messages, AM-Hama outperforms Hama by ratios of thousands. GraphHP nonetheless manages to further reduce them by ratios of tens. On execution time, AM-Hama outperforms Hama by ratios of around 2. Compared with AM-Hama,

GraphHP reduces the time by ratios of tens (from 600+ to 50+). These experimental results demonstrate the huge performance advantage of GraphHP over Hama and AM-Hama. For GraphHP, as the number of partitions increases, the required number of iterations increases only marginally while the number of network messages increases modestly. These observations bode well for its scalability. GraphHP can save huge numbers of global iterations and network messages. But it may consumes more time on average for each global iteration because of the iterative execution of pseudo-superstep. As a result, the performance advantage of GraphHP measured by time is not as significant as those measured by iterations and messages.

| Platforms | I | M(mil) | T(sec) |
|---|---|---|---|
| Hama | 10671 | 43,829 | 17912 |
| AM-Hama | 10593 | 387 | 5792 |
| GraphHP | 451 | 71 | 2155 |

**Table 2: SSP Evaluation Results on USA-Road-Full**

The experimental results on the bigger USA-Road-Full dataset are similar. The detailed results for the case of 108 partitions are presented in Table 2, in which I, M and T represent number of iterations, number of network messages and execution time respectively. Compared with Hama, AM-Hama saves huge number of network messages but only reduces the number of iterations by small margins. GraphHP performs significantly better than both of them.

### 7.3 PageRank

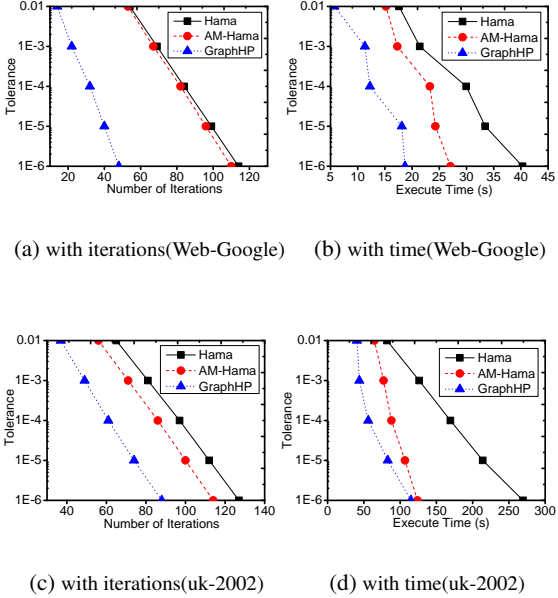

**Figure 4: Convergence of PageRank**

On the PageRank computation, we evaluate the performance of the incremental BSP algorithm presented in Algorithm 5 on the two datasets. We first compare the convergence rates of PageRank in terms of number of iterations and execution time on three platforms. In the preprocessing step, Web-Google and uk-2002 datasets are divided into 12 and 72 partitions respectively. The tolerance threshold $\Delta$ is set from *1E-2* to *1E-6*. The comparative results on Web-Google are shown in Figure 4 (a) and (b). Given a tolerance threshold, GraphHP requires considerably less iterations than Hama. Moreover, as the tolerance threshold becomes smaller, the number of required iterations increases more rapidly on Hama than on GraphHP. A similar trend is observed in the convergence rate with execution time. GraphHP consumes considerably less time than Hama. When the tolerance threshold decreases, its time also increases more slowly compared with Hama. It can also be observed that AM-Hama performs better than Hama, but still considerably worse than GraphHP. Even though AM-Hama can only slightly reduce the number of required iterations compared with Hama, it outperforms Hama by considerable margins in term of time. The asynchronous messaging mechanism significantly reduces the number of network messages, thus improving the overall performance of AM-Hama. The evaluations results on the uk-2002 dataset, as shown in Figure 4 (c) and (d), follow a similar comparative pattern. GraphHP consistently performs better than both Hama and AM-Hama.

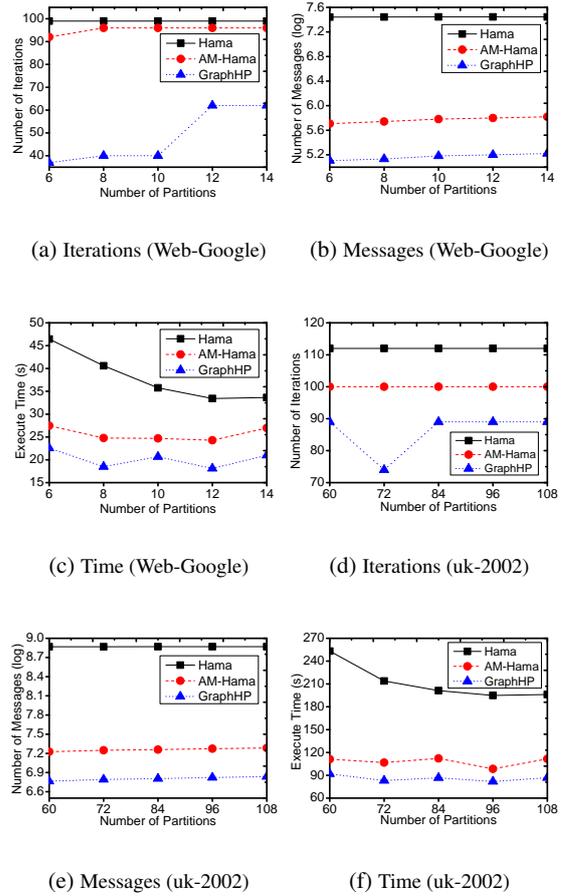

**Figure 5: Scalability Evaluation of PageRank**

We then study scalability of the platforms by measuring their performance variations as input graphs are divided into increasing numbers of partitions. The tolerance threshold $\Delta$ is set to be *1E-4*. The evaluation results, as shown in Figure 5, are similar on two datasets. We only present the results up to 14 and 108 partitions on the two datasets respectively because after that, further partitioning can not improve the parallel performance. Note that because of huge performance difference, the numbers of network messages in the figure are presented by a *logarithmic* scale (base 10). Regardless of the number of partitions, GraphHP consistently outforms Hama and AM-Hama by considerable margins in term of every metric. Since asynchronous messaging mechanism effectively reduces the numbers of iterations and network messages on both datasets, AM-Hama performs much better than Hama. On the other hand, the superior performance of GraphHP over AM-Hama demonstrates that the local pseudo-superstep iteration of GraphHP manages to further reduce both numbers, optimizing the synchronization and communication overhead. It is worthy to point out that as the number of partitions increases, the numbers of required iterations and network messages increase only slightly on GraphHP. This observation bodes well for its scalability.

## 7.4 Bipartite Matching

| Platforms | cit-patents | | | delaunay_n24 | | |
|---|---|---|---|---|---|---|
| | I | M(mil) | T(s) | I | M(mil) | T(s) |
| Hama | 23 | 41.5 | 42.9 | 15 | 126.64 | 83.3 |
| AM-Hama | 20 | 4.4 | 21.6 | 15 | 0.16 | 34.9 |
| GraphHP | 7 | 3.0 | 13.0 | 5 | 0.10 | 15.9 |

**Table 3: BM Evaluation Results**

For the bipartite matching algorithm, all the platforms requires only a small number of iterations. We therefore do not provide with the detailed performance variation of the platforms with the number of partitions. We instead present a snapshot of their performance on the two real datasets in Table 3. The two datasets, cit-patent and delaunay_n24, are divided into 18 and 48 partitions respectively. On the cit-patent dataset, Hama requires only 20+ iterations. GraphHP nonetheless manages to reduce the number of required iterations by a ratio of more than 3 (from 23 to 7). It also reduces the execution time by a ratio of more than 3 (from 42s to 13s). Compared with Hama, AM-Hama manages to significantly reduce the communication overhead, but takes a only slightly less number of iterations. It can be observed that GraphHP also outperforms AM-Hama by considerable margins on every metric.

The evaluation results on the delaunay_n24 dataset are similar. All the platforms iteratively executes the 3-way handshake to establish maximal matching. By executing the handshakes within a partition in a single iteration, GraphHP effectively reduces the number of iterations. As a result, it performs considerably better than both Hama and AM-Hama.

## 7.5 Comparison with Giraph++ and GraphLab

To illustrate the potential advantage of GraphHP over Giraph++ and GraphLab, we compare their performance for the PageRank algorithm on the Web-Google dataset. Suppose that the input graph is divided into 12 partitions. All the platforms are run on the cluster with the same hardware setting described in Section 7.1.

| Platforms | Tolerance(1E-3) | | | Tolerance(1E-4) | | |
|---|---|---|---|---|---|---|
| | I | M(k) | T(s) | I | M(k) | T(s) |
| GraphLab(Sync) | 92 | – | 43.0 | 106 | – | 54.8 |
| GraphLab(Async) | – | – | 82.4 | – | – | 106.4 |
| Giraph++ | 46 | 450 | 13.9 | 54 | 600 | 17.6 |
| GraphHP | 32 | 125 | 11.2 | 40 | 158 | 14.7 |

**Table 4: Comparing GraphHP with Giraph++ and GraphLab**

Since Giraph++ is only publicly available on Giraph, we implement an improvised version of the PageRank algorithm for Giraph++, whose pseudo-code is presented in [32], by writing a corresponding `bsp()` function on Hama. The `bsp()` function specifies the sequential PageRank computations at each partition at each superstep iteration. The comparative results between GraphHP and Giraph++ are presented in Table 4. On Giraph++, the PageRank implementation sequentially update each vertex once and immediately propagates its update to its neighboring vertices within a same partition. On GraphHP, the PageRank values of the vertices within a partition are iteratively updated until they converge. Therefore, GraphHP requires less global iterations than Giraph++. As the global iteration proceeds, the number of vertices that needs to propagate their PageRank updates across partitions decreases more rapidly on GraphHP than on Giraph++. These two factors combined result in the significantly reduced number of network messages on GraphHP. These experimental results clearly demonstrate the potential advantage of GraphHP over Giraph++ in optimizing the synchronization and communication overhead.

The comparative evaluation results between GraphHP and GraphLab (version 2.2) are also presented in Table 4. GraphLab has two processing modes, *Sync* and *Async*. The *Sync* mode uses an iteration mechanism similar to the superstep iteration of the standard BSP execution model to repeatedly update the PageRank values of vertices until they converge. It takes even more iterations than Hama. Even though Async often needs fewer iterations to converge, it also reduces the degree of parallelism due to the locking mechanism to ensure data consistency. For the PageRank computation on the test dataset, this trade-off results in the inferior performance of Async. GraphHP requires much less iterations than GraphLab Sync and outperforms both modes by considerable margins.

## 8. RELATED WORK

In Section 1 and 2, we analyzed the main work on the MapReduce and BSP frameworks for large-scale graph processing. In this section, we continue to review more related work in the platforms for distributed graph processing.

Even before Google's Pregel, there have been several general BSP library implementations, such as Green BSP Library [15] and BSPlib [17]. However, these systems do not provide a graph-specific API, not to mention a vertex-centric programming interface. Their scalability and fault tolerance have neither been evaluated on large-scale clusters. The platforms like Parallel BGL [16] and CGM-graph [9] provide graph-specific implementation API based on MPI. But they do not provide vertex-centric programming interfaces. Neither do they address the critical fault tolerance issue.

Besides GraphLab, there are several asynchronous abstractions [25, 26] that can also facilitate parallel graph algorithm implementations. However, these systems do not ensure serializability, or do not provide adequate mechanisms to recover from data races. Grace [35] is also an asynchronous graph processing platform. It is built on a single machine and employs the similar vertex-centric programming interface as BSP platforms, but uses customization of vertex scheduling and message selection to support asynchronous computation. Even though these asynchronous platforms can accelerate convergence computations within parts of an input graph, the trade-off is that they require heavy scheduling overhead.

Other miscellaneous platforms include Trinity [29] and Kineograph [11]. Trinity stores graph data in a distributed memory to support online graph processing. For offline processing, it uses an execution model similar to that of typical BSP platforms. Kineograph [6] is a distributed system for storing continuously changing graphs. Even though its computation model is vertex centric, graph mining algorithms are still performed on static snapshots of changing graphs.

## 9. CONCLUSION AND FUTURE WORK

Implementing efficient graph algorithms on BSP platforms can be challenging due to high synchronization and communication overhead. In this paper, we proposed a hybrid execution model to optimize synchronization and communication overhead. It achieves the goal by executing a sequence of pseudo-supersteps for local computations at each global iteration. We have built a corresponding hybrid platform GraphHP based on Hama, demonstrating how the hybrid execution model can be easily implemented within the BSP abstraction. Our comprehensive experiments have also validated the efficacy of the hybrid approach.

Future work can be pursued on multiple fronts. Because of pseudo-superstep iteration, GraphHP may consume considerable time on a global iteration. It is interesting to investigate how to speed up the execution of pseudo-superstep iteration without sacrificing the uniform vertex-centric programming interface. On the other hand, load balancing is important for efficient BSP processing. Existing BSP load balancing techniques [10, 21] are for the standard execution engine. It is challenging to design an effective load balancing approach applicable to GraphHP.

## 10. REFERENCES

[1] Apache Giraph. http://giraph.apache.org/.
[2] Apache Hadoop. http://hadoop.apache.org/.
[3] Apache Hama. http://hama.apache.org/.
[4] The University of Florida sparse matrix collection. http://www.cise.ufl.edu/research/sparse/matrices/.
[5] USA road network. http://www.dis.uniroma1.it/challenge9/download.shtml.
[6] T. Anderson, S. Owicki, J. Saxe, and C. Thacker. High-speed switch scheduling for local-area networks. *ACM Trans. Comp. Syst.*, 11(4):319–352, 1993.
[7] N. T. Bao and T. Suzumura. Towards highly scalable pregel-based graph processing platform with x10. In *WWW*, pages 501–507, 2013.
[8] S. Brin and L. Page. The anatomy of a large-scale hypertextual web search engine. In *WWW*, pages 107–117, 1998.
[9] A. Chan and F. Dehne. CGMGRAPH/CGMLIB: Implementing and testing CGM graph algorithms on PC clusters and shared memory machines. *Intl. J. of High Performance Computing Applications*, 19(1):81–97, 2005.
[10] R. Chen, X. Weng, B. He, M. Yang, B. Choi, and X. Li. Improving large graph processing on partitioned graphs in the cloud. In *SOCC*, 2012.
[11] R. Cheng, J. Hong, A. Kyrola, Y. Miao, X. Weng, M. Wu, F. Yang, L. Zhou, F. Zhao, and E. Chen. Kineograph: taking the pulse of a fast-changing and connected world. In *EuroSys*, pages 85–98, 2012.
[12] B. V. Cherkassky, A. V. Goldberg, and T. Radzik. Shortest paths algorithms: Theory and experimental evaluation. *Mathematical Programming*, 73:129–174, 1996.
[13] J. Dean and S. Ghemawat. MapReduce: simplified data processing on large clusters. In *OSDI*, 2004.
[14] J. Gonzalez, Y. Low, and C. Guestrin. Residual splash for optimally parallelizing belief propagation. In *AISTATS*, pages 177–184, 2009.
[15] M. W. Goudreau, K. Lang, S. B. Rao, T. Suel, and T. Tsantilas. Portable and efficient parallel computing using the bsp model. *IEEE Trans. Comp.*, 48(7):670–689, 1999.
[16] D. Gregor and A. Lumsdaine. BGL: A generic library for distributed graph computations. In *Proc. of Parallel Object-Oriented Scientific Computing (POOSC)*, 2005.
[17] J. Hill, B. McColl, D. Stefanescu, M. Goudreau, K. Lang, S. Rao, T. Suel, T. Tsantilas, and R. Bisseling. BSPlib: The BSP programming library. *Parallel Computing*, 24:1947–1980, 1998.
[18] M. Isard, M. Budiu, Y. Yu, A. Birrell, and D. Fetterly. Dryad: distributed data-parallel programs from sequential building blocks. In *EuroSys*, 2007.
[19] G. Karypis and V. Kumar. A coarse-grain parallel formulation of multilevel k-way graph partitioning algorithm. In *SIAM PP*, 1997.
[20] G. Karypis and V. Kumar. A fast and high quality multilevel scheme for partitioning irregular graphs. *SIAM J. Sci. Comput.*, 20(1):359–392, 1998.
[21] Z. Khayyat, K. Awara, A. Alonazi, H. Jamjoom, D. Williams, and P. Kalnis. Mizan: A system for dynamic load balancing in large-scale graph processing. In *EuroSys*, 2013.
[22] Y. Low, D. Bickson, J. Gonzalez, C. Guestrin, A. Kyrola, and J. M. Hellerstein. Distributed GraphLab: a framework for machine learning and data mining in the cloud. *PVLDB*, 5(8):716–727, 2012.
[23] G. Malewicz, M. H. Austern, A. J. Bik, J. Dehnert, I. Horn, N. Leiser, and G. Czajkowski. Pregel: a system for large-scale graph processing. In *SIGMOD*, pages 135–146, 2010.
[24] R. Neal and G. Hinton. A view of the em algorithm that justifies incremental, sparse, and other variants. In *Learning in Graph Models*, pages 355–389, 1998.
[25] R. Pearce, M. Gokhale, and N. Amato. Multithreaded asynchronous graph traversal for in-memory and semi-external memory. In *SC*, pages 1–11, 2010.
[26] R. Power and J. Li. Piccolo: building fast, distributed programs with partitioned tables. In *OSDI*, 2010.
[27] S. Salihoglu and J. Widom. GPS: A graph processing system. In *SSDBM*, 2013.
[28] S. Salihoglu and J. Widom. Optimizing graph algorithms on pregel-like systems. Technical report, Stanford InfoLab, 2013.
[29] B. Shao, H. Wang, and Y. Li. Trinity: A distributed graph engine on a memory cloud. In *SIGMOD*, pages 505–516, 2013.
[30] A. J. Smola and S. Narayanamurthy. An architecture for parallel topic models. *PVLDB*, 3(1):703–710, 2010.
[31] S.Wasserman and K.Faust. *Social Network Analysis*. Cambridge University Press, 1994.
[32] Y. Tian, A. Balmin, S. A. Corsten, S. Tatikonda, and J. McPherson. From "think like a vertex" to "think like a graph". *PVLDB*, 7(3), 2013.
[33] T.Washio and H.Motoda. State of the art of graph-based data mining. *SIGKDD Explorations Newsletter*, 5(1):59–68, 2003.
[34] L. G. Valiant. A bridging model for parallel computation. *Comm. ACM*, 33(8):103–111, 1990.
[35] G. Wang, W. Xie, A. Demers, and J. Gehrke. Asynchronous large-scale graph processing made easy. In *CIDR*, 2013.
[36] Y. Zhang, Q. Gao, L. Gao, and C. Wang. Accelerate large-scale iterative computation through asynchronous accumulative updates. In *ScienceCloud*, pages 13–22, 2012.